\documentclass[10pt,aps,prl,twocolumn,notitlepage,superscriptaddress,preprintnumbers]{revtex4-1}

\usepackage{amsmath,amssymb,amsfonts}
\usepackage{bm}
\usepackage{graphicx,color}
\usepackage{verbatim}
\usepackage{float}
\usepackage{dcolumn}
\usepackage{natbib}
\usepackage{hyperref}

\newcommand{\hA}{\mathsf A}
\newcommand{\hW}{\mathsf W}

\newcommand{\bx}{{\bf x}}

\newcommand{\be}{\begin{equation}}
\newcommand{\ee}{\end{equation}}
\newcommand{\bea}{\begin{eqnarray}}
\newcommand{\eea}{\end{eqnarray}}
\newcommand{\ie}{\emph{i.e.}}
\newcommand{\eg}{\emph{e.g.}}

\newcommand{\ER}{Erd\"{o}s-R\'enyi }
\newcommand{\avg}[1]{\langle #1\rangle}

\begin{document}

\title{The Grand Canonical Ensemble of Weighted Networks}

\author{Andrea Gabrielli}
\affiliation{Istituto dei Sistemi Complessi (CNR) UoS Sapienza, P.le A. Moro 2, 00185 Rome (Italy)}
\affiliation{IMT School for Advanced Studies, Piazza San Francesco 19, 55100 Lucca (Italy)}
\author{Rossana Mastrandrea}\email{rossana.mastrandrea@imtlucca.it }
\affiliation{IMT School for Advanced Studies, Piazza San Francesco 19, 55100 Lucca (Italy)}
\author{Guido Caldarelli}
\affiliation{IMT School for Advanced Studies, Piazza San Francesco 19, 55100 Lucca (Italy)}
\affiliation{Istituto dei Sistemi Complessi (CNR) UoS Sapienza, P.le A. Moro 2, 00185 Rome (Italy)}
\author{Giulio Cimini}
\affiliation{IMT School for Advanced Studies, Piazza San Francesco 19, 55100 Lucca (Italy)}
\affiliation{Istituto dei Sistemi Complessi (CNR) UoS Sapienza, P.le A. Moro 2, 00185 Rome (Italy)}

\begin{abstract}
The cornerstone of statistical mechanics of complex networks is the idea that the links, and not the nodes, are the effective particles of the system. 
Here we formulate a mapping between weighted networks and lattice gasses, making the conceptual step forward of interpreting weighted links as particles with a generalized coordinate. 
This leads to the definition of the grand canonical ensemble of weighted complex networks. 
We derive exact expressions for the partition function and thermodynamic quantities, 
both in the cases of global and local (\ie, node-specific) constraints on density and mean energy of particles. 
We further show that, when modeling real cases of networks, the binary and weighted statistics of the ensemble can be disentangled, 
leading to a simplified framework for a range of practical applications.
\end{abstract}

\maketitle


What distinguishes a network from the systems typically studied in physics is the complex heterogeneous pattern of interactions (links) among its constituent elements (nodes). 
Indeed, the statistical mechanics approach to networks has been developed treating the interactions themselves as the degrees of freedom of the system, 
pushing forward the interpretation of links as the actual particles of the system \cite{cimini2018review}. 
Under this view, the maximum number of particles $V$ (\ie, the maximum number of links) is the equivalent of the volume of a physical system.
For binary networks with fixed number $N$ of nodes, a number of seminal works \cite{dorogovtsev2003principles,park2004statistical,garlaschelli2006multispecies} 
defined the canonical ensemble by fixing the number of links $L$, and the grand canonical ensemble by letting $L$ fluctuate around its expected value. 
For instance in the \ER model, these two cases correspond to $G(L)$ and $G(p)$ with $p=L/V$ denoting the link probability. 
The microcanonical ensemble is retrieved in this framework upon defining an energy function of network configurations, which however 
unlike in physical systems cannot be derived from first principles \cite{palla2004statisitcal,bogacz2006homogeneous}. 
This difficulty led statistical mechanics of networks to be re-framed more closely to information theory, according to Jayne's formulation \cite{jaynes1957information}. 
Indeed, nowadays the microcanonical ensemble is defined by assigning equal probability to the network configurations that satisfy a given set of structural constraints exactly, 
whereas, in the canonical ensembles network probabilities are such that the constraints are met on average over the ensemble 
\cite{bianconi2008entropy,squartini2011analytical}. 
Notably, this framework naturally incorporates networks with weighted interactions \cite{barrat2004architecture}, by treating links as multiple particle states. 
In particular, the canonical ensemble has been derived for networks with links assuming integer weights \cite{garlaschelli2009generalized,mastrandrea2014enhanced} 
and approximately for networks with (distinguishable) multilinks \cite{sagarra2013statistical,sagarra2015role}. 

In this work we push forward this idea of considering links as particles by assuming the weights of existing links 
to be generalized coordinates (\eg, energy or magnetic moment) associated to such particles \footnote{A different analogy between evolving networks and equilibrium Bose gases 
consists in treating nodes as energy levels and links as non-interacting particles \cite{bianconi2001bose}}. 
This allows defining the proper statistical mechanics formulation of the grand canonical ensemble of networks---in strict analogy with the case of lattice gases. 
Here for simplicity we focus our discussion on undirected networks with links assuming continuous weights. 

\begin{figure}
\centering
\includegraphics[width=7cm]{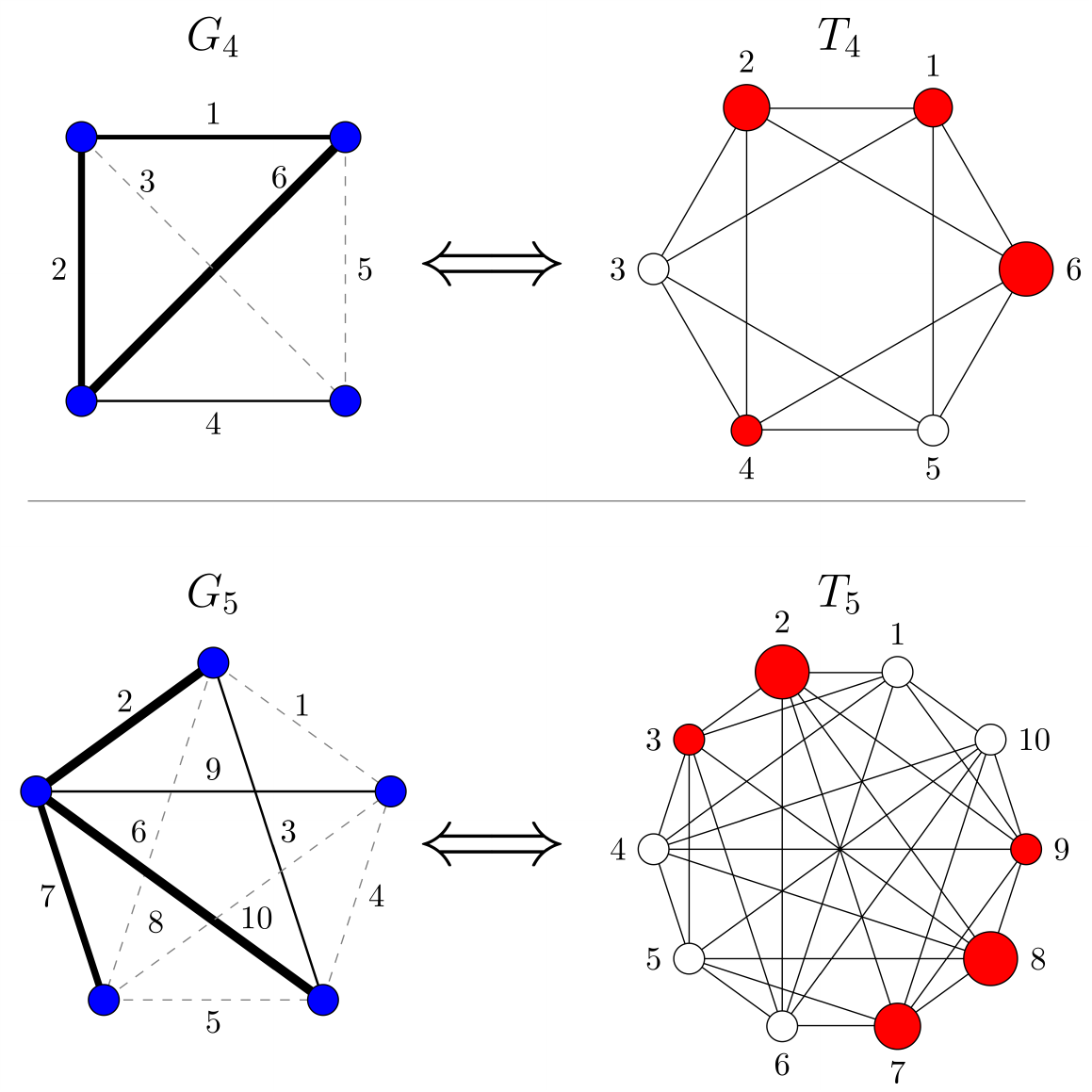}
\caption{Mapping between an undirected graph $G_N$ and a lattice gas on the corresponding triangular graph $T_N$ of $K_N$ (we report the illustrative examples $N=4,5$). 
Only the existing links of $G_N$ (represented as solid black lines) are placed as particles on the lattice sites (represented as full dots), 
and the weight of such links (given by lines' thickness) corresponds to the generalized coordinates of particles (given by dots' size). 
Note that two particles in the lattice gas are neighbors if the corresponding links in $G_N$ have a node in common.}\label{fig:Tfig}
\end{figure}

We define a mapping between a simple undirected weighted network $G$ with $N$ nodes and a lattice gas as follows. 
First we note that each link of $G$ corresponds to an edge of $K$, the complete simple graph of $N$ nodes. 
Thus we define a lattice using the {\em line graph} of $K$, also called in this case the {\em triangular graph} $T$ of order $N$ \cite{brualdi1991}. 
This is the graph obtained by associating a vertex with each edge of $K$, and connecting two vertices with an edge iff the corresponding edges of $K$ have a vertex in common. 
Given this representation, we map each link of $G$ with weight $w$ into a particle with internal coordinate $w$ occupying the corresponding vertex (lattice site) of the graph $T$ (see Fig. \ref{fig:Tfig}). 
Therefore we have that the binary adjacency matrix $\hA=\{a_{ij}\}_{(i,j)\in{\cal V}}$ of the network fixes the positions of the gas particles on the lattice $T$, 
whereas, its weighted adjacency matrix $\hW=\{w_{ij}\}_{(i,j)\in{\cal V}}$ defines the internal coordinates of existing particles. 
Here ${\cal V}$ denotes the set of all unordered node pairs, with $|{\cal V}|=V=N(N-1)/2$ being the volume of the system.  

This mapping allows formulating in rigorous way the Grand Canonical ensemble of complex weighted networks. 
We first define the configuration ${\cal C}$ of an undirected weighted network as the pair $(\hA,\hW)$: 
the set of existing links $(i,j)\in {\cal L}\subseteq{\cal V}$ with $|{\cal L}|=L$ (\ie, the set of node pairs with $a_{ij}=1$)
and the set of weights $\{w_{ij}\}_{(i,j)\in {\cal L}}$ associated to them (meaning that only existing links/particles contribute to the statistics of the system). 
Therefore, the grand canonical probability distribution is $P({\cal C})=P(\hA, \hW)\equiv P({\cal L}, \{w_{ij}\}_{(i,j)\in {\cal L}})$, 
and the sum over configuration is performed as
\be
\sum_{\cal C}\equiv \sum_{\hA} \prod_{i<j}^{{\cal L}} \int_{0}^{\infty} dw_{ij}.
\label{sum-WN}
\ee
The information entropy associated to the probability measure $P(\cal C)$ is as usual $S=-\sum_{\cal C}P(\cal C)\log P(\cal C)$, 
and the shape of $P(\cal C)$ is found by maximizing $S$ under given constraints. 
This is the framework of Exponential Random Graph models \cite{park2004statistical,garlaschelli2009generalized,desmarais2012statistical}.

{\em Global constraints} --- The simplest nontrivial ensemble of this kind is obtained by imposing the mean total number of links (particles) $\avg{L}\equiv\avg{\sum_{i<j}^{{\cal V}}a_{ij}}=L^*$ 
and the mean total weight (\eg, energy) $\avg{W}\equiv\avg{\sum_{i<j}^{{\cal L}}w_{ij}}=W^*$, where the average is defined by the measure $P(\hA,\hW)$. 
This is the weighted version of the \ER model. We get $P(\hA,\hW,\alpha,\beta)=Z_G^{-1}(\alpha,\beta)e^{-H(\hA,\hW,\alpha,\beta)}$ with the Hamiltonian
\be
H(\hA,\hW,\alpha,\beta)=\alpha\sum_{i<j}^{{\cal V}}a_{ij}+\beta\sum_{i<j}^{{\cal L}}w_{ij},
\label{GCE-WER}
\ee
where $\alpha$ and $\beta$ are the Lagrange multipliers related to $L$ and $W$ respectively. 
$Z_G(\alpha,\beta)$ is, in analogy with statistical mechanics, the grand canonical partition function 
\be
Z_G(\alpha,\beta)=\sum_{\cal C}e^{-H(\hA,\hW,\alpha,\beta)}=\sum_{\hA}e^{-\alpha\sum_{i<j}^{\cal V}a_{ij}}Z_C(\beta)
\label{GCPF-WER}
\ee
where $Z_C(\beta)=\prod_{i<j}^{{\cal L}}\int_0^\infty dw_{ij}e^{-\beta w_{ij}}=\beta^{-\sum_{i<j}^{\cal V}a_{ij}}=\beta^{-L}$ is the canonical partition function. 
The sum in Eq.~\eqref{GCPF-WER} is easily performed by noting that 
$\sum_{\hA}e^{-\alpha \sum_{i<j}^{\cal V}a_{ij}}\beta^{-\sum_{i<j}^{\cal V}a_{ij}}=\sum_{L=0}^{V} n_{\cal C}(L)e^{-\alpha L}\beta^{-L}$ 
where $n_{\cal C}(L)={V \choose L}$ is the number of binary configurations with exactly $L$ links. We finally have
\be
Z_G(\alpha,\beta)=\sum_{L=0}^{V}{V \choose L} 
\frac{e^{-\alpha 
L}}{\beta^L}=\left[1+\frac{e^{-\alpha}}{\beta}\right]^V\,.
\ee
The equations determining the values of $\alpha$ and $\beta$ are then
\bea
&&\avg{L}\equiv -\partial_{\alpha} \log Z_G(\alpha,\beta)\equiv \frac{V}{\beta e^\alpha+1}=L^*,\\
&&\avg{W}\equiv-\partial_{\beta} \log Z_G(\alpha,\beta)\equiv \frac{V\beta^{-1}}{\beta e^\alpha+1}=W^*,
\eea
from which we immediately find $\beta^{-1}=W^*/L^*=w^*$, \ie, the mean weight, and $1+e^\alpha/w^*=V/L^*$. 
We thus see that while $\beta$ controls for the mean weight (energy) of existing links (particles), $\alpha$ controls for the mean density of links (particles). 
Note that since between each pair of nodes there can be only a single link/particle, the system can be represented 
with $V$ copies of a Fermi system having a single energy level $\varepsilon=1$. 
Under this analogy, $\log\beta$ plays the role of the inverse absolute temperature $(kT)^{-1}$, 
whereas, $-\alpha$ is the ratio $\mu(kT)^{-1}$ between chemical potential and temperature. 

Remarkably, we can perform the parameter transformation $\alpha'=\alpha+\log\beta$, 
so that $\alpha'$ alone determines the mean link density and, given this density, $\beta$ 
alone sets the mean weight of existing links: we have 
\be
P(\hA,\hW)=\left[\prod_{i<j}^{\cal V}\frac{e^{-\alpha' a_{ij}}}{1+e^{-\alpha'}}\right]\left[\prod_{i<j}^{{\cal L}}\beta e^{-\beta w_{ij}}\right].
\label{PAW}
\ee
This shows that individual link occupations are all mutually independent 
events and that, given a binary configuration $\hA$, weight values of individual existing links 
are also independent events. Besides, moments of link occupation 
and of link weight probability distributions can be independently set in order 
to satisfy the constraints. As explicitly shown in the next section, this property is due to the global nature of the constraints. 
Note that as for equilibrium statistical mechanics with short range interactions, if the system is homogeneous then local and global measures coincide.

{\em Local constraints} --- We now impose for each node $i$ the mean {\em degree} or number of incident links $\avg{k_i}\equiv\avg{\sum_{j(\ne i)}^{\cal V} a_{ij}}=k_i^*$ 
and the mean {\em strength} or total weight of incident links $\avg{s_i}=\avg{\sum_{j(\ne i)}^{\cal L} w_{ij}}=s_i^*$. 
This grand canonical ensemble is analogous to the continuous version of the {\em enhanced configuration model} \cite{mastrandrea2014enhanced}, whence we use the acronym CECM. 
We have $P(\hA,\hW,\{\alpha_i,\beta_i\}_{i=1}^N)=Z_G^{-1}\left(\{\alpha_i,\beta_i\}_{i=1}^N\right)e^{-H(\hA,\hW,\{\alpha_i,\beta_i\}_{i=1}^N)}$ with
\be
H(\hA,\hW,\{\alpha_i,\beta_i\}_{i=1}^N)=\sum_{i<j}^{\cal V}(\alpha_i+\alpha_j)a_{ij}+\sum_{i<j}^{{\cal L}}(\beta_i+\beta_j)w_{ij},
\label{GCE-CM}
\ee
\bea
Z_G\left(\{\alpha_i,\beta_i\}_{i=1}^N\right)&=&\sum_{\cal C}e^{-H(\hA,\hW,\{\alpha_i,\beta_i\}_{i=1}^N)}=\nonumber\\
&=&\sum_{\hA}e^{-\sum_{i<j}^{\cal V}(\alpha_i+\alpha_j)a_{ij}}Z_C\left(\{\beta\}_{i=1}^N\right),
\label{GCPF-CM}
\eea
and $Z_C\left(\{\beta\}_{i=1}^N\right)=\prod_{i<j}^{{\cal L}}(\beta_i+\beta_j)^{-1}$. 
Performing the sum over all binary configurations leads to
\begin{widetext}
\be
Z_G\left(\{\alpha_i,\beta_i\}_{i=1}^N\right)=\sum_{\hA}\frac{e^{-\sum_{i<j}^{\cal V}(\alpha_i+\alpha_j)a_{ij}}}{\prod_{i<j}^{{\cal L}}(\beta_i+\beta_j)}
=\sum_{\hA}\prod_{i<j}^{{\cal L}}\frac{e^{-(\alpha_i+\alpha_j)a_{ij}}}{\beta_i+\beta_j}=1+\sum_{\cal U\subset V}\prod_{i<j}^{\cal U} 
\frac{e^{-(\alpha_i+\alpha_j)}}{\beta_i+\beta_j}=\prod_{i<j}^{\cal V}\left(1+\frac{e^{-(\alpha_i+\alpha_j)}}{\beta_i+\beta_j}\right)
\label{GCPF-CM2}
\ee
where ${\cal U}$ is a generic subset of ${\cal V}$. 
The values of the multipliers are then found through the constraints equations:
\bea
\avg{k_i}&\equiv&-\partial_{\alpha_i}\log Z_G\left(\{\alpha_l,\beta_l\}_{l=1}^N\right)\equiv \sum_{j(\ne i)}^{\cal V}\frac{1}{1+(\beta_i+\beta_j)e^{\alpha_i+\alpha_j}}=k_i^*,\label{constr-GCPF-CM1}\\
\avg{s_i}&\equiv&-\partial_{\beta_i}\log Z_G\left(\{\alpha_l,\beta_l\}_{l=1}^N\right)\equiv \sum_{j(\ne i)}^{\cal V}\frac{(\beta_i+\beta_j)^{-1}}{1+(\beta_i+\beta_j)e^{\alpha_i+\alpha_j}}=s_i^*,\label{constr-GCPF-CM2}
\eea
$\forall i$. Note that after some algebra we can rewrite $P(\hA,\hW)$ as 
\be
P(\hA,\hW)=\left[\prod_{i<j}^{\cal V}\frac{e^{-[\alpha_i+\alpha_j+\log (\beta_i+\beta_j)]a_{ij}}}{1+e^{-[\alpha_i+\alpha_j+\log (\beta_i+\beta_j)]}}\right]
\left[\prod_{i<j}^{{\cal L}}(\beta_i+\beta_j)e^{-(\beta_i+\beta_j)w_{ij}}\right]=\pi(\hA)q(W_{{\cal L}})
\label{GCPF-CM3}
\ee
\end{widetext}
with $\pi(\hA)$ being the unconditional probability distribution of the binary configuration $\hA$, 
and $q(W_{{\cal L}})$ the probability density function of the weights of the existing links (\ie, the set ${\cal L}$) conditional to $\hA$.
The form of $q(W_{{\cal L}})$ is exponential, differently from the geometric and Poissonian forms obtained in 
\cite{mastrandrea2014enhanced} and \cite{sagarra2015role} respectively, due to the continuous nature of the weights.

At this point some considerations are in order.
I) Both $\pi(\hA)$ and $q(W_{{\cal L}})$ factorize into the product of single 
link probability distributions: occupations of different links are 
independent events and, conditional to the binary configuration, weights of 
different links are also independent. 
II) However the parameters defining single link probabilities and weights are entangled, 
which means that local link densities cannot be set independently on local weights, because of the simultaneous conservations of mean node degrees and strengths. 
Such an interplay allows better clarifying the role of nodes (and in particular of node heterogeneity), which play the role of interactions between links (particles). 
Indeed only if node properties are homogeneous, like when we impose global constraints, such topological interactions disappear: 
the system is spatially homogeneous in terms of density of particles and of energy, which can be thus set independently. 
The statistical mechanical case analogous to a heterogeneous network situation instead arises when we 
constrain the local mean particle and energy densities, $n(\bx)$ and $\varepsilon(\bx)$, to be heterogeneous, \ie, both dependent on $\bx$. 
This case is typically not encountered in ordinary equilibrium statistical mechanics, with the possible exception of glassy disordered systems and long range interactions.  
III) If we look at the generic link occupation probability
\be
p_{ij}=\frac{e^{-[\alpha_i+\alpha_j+\log (\beta_i+\beta_j)]}}{1+e^{-[\alpha_i+\alpha_j+\log (\beta_i+\beta_j)]}}
\label{ecm_prob}
\ee
from the viewpoint of statistical mechanics, we can again interpret the single link problem as a single state local Fermi system with energy level $\varepsilon=1$, 
inverse local temperature $(kT_{ij})^{-1}=\log (\beta_i+\beta_j)$, and local chemical potential $\mu_{ij}=-kT_{ij}(\alpha_i+\alpha_j)$. 
However, differently from the homogeneous case, different links are not independent copies of the same problem, 
but topologically interacting single-level Fermi systems with different local temperatures and chemical potentials---which are mutually related by local heterogeneous constraints.

\begin{figure*}
\centering
\includegraphics[width=17.2cm]{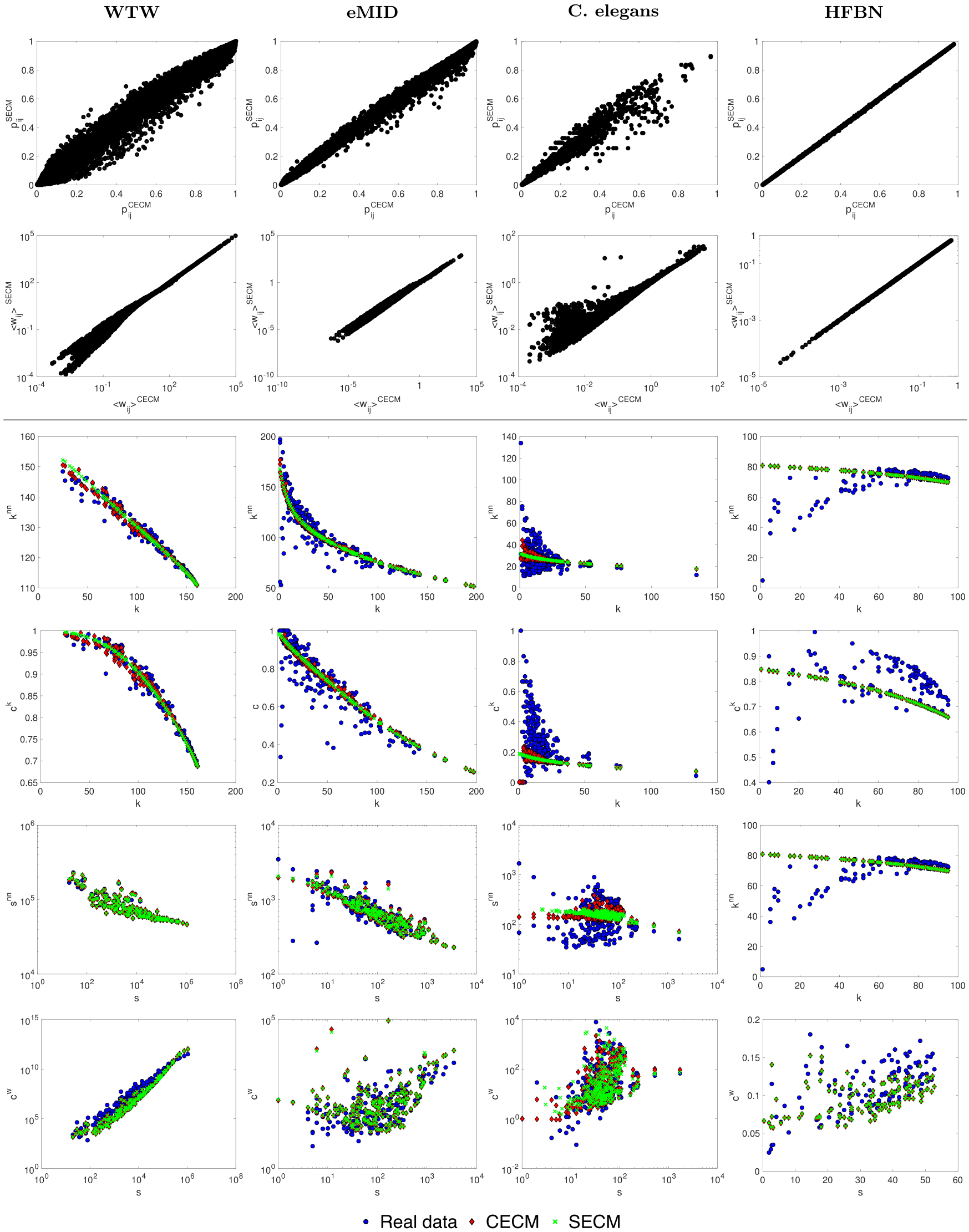}
\caption{Properties of CECM and SECM ensembles in four real networks: the World Trade Web~\cite{mastrandrea2014enhanced}, the eMID interbank network~\cite{mastrandrea2014enhanced}, 
the neural network of C. elegans \cite{oshio2003database} and the human functional brain network (HFBN) \cite{mastrandrea2017organization}. 
The upper part of the figure shows the comparison of link probabilities (first row) and of expected weights (second row) obtained by CECM and SECM.
The lower part of the figure instead shows how the two ensembles reproduce higher-order statistics of the real networks (defined in \cite{cimini2015estimating}): 
nearest neighbors degree $k^{nn}$ (third row), clustering coefficient $c$ (fourth row), nearest neighbors strength $s^{nn}$ (fifth row) and weighted clustering $c^w$ (sixth row).} 
\label{fig:CECM}
\end{figure*}

{\em Separability of binary and weighted statistics} --- We finally explore the separability of local links densities and weights distribution 
also for the case of local constraints \cite{almog2015gdp}. To this end we introduce a two-step entropy maximization procedure, the {\em separable} enhanced configuration model (SECM):
\begin{enumerate}
\item We first constrain the mean node degrees only, obtaining the probability of the binary configuration $\hA$ as for the standard configuration model \cite{park2004statistical}:
\be
\pi(\hA)=\prod_{i<j}^{\cal V}\frac{e^{-(\alpha'_i+\alpha'_j) a_{ij}}}{1+e^{-(\alpha'_i+\alpha'_j)}}\,,
\label{P-A-TS}
\ee
\item Then, for each $\hA$, we constrain the mean node strengths, obtaining the probability density of the link weights conditional to $\hA$ (coinciding with that of the CECM):
\be
q(W_{{\cal L}})=\prod_{i<j}^{{\cal L}}(\beta_i+\beta_j)e^{-(\beta_i+\beta_j)w_{ij}}.
\ee
\end{enumerate}
The SECM is thus defined by the constraint equations
\bea
\avg{k_i}&&\equiv \sum_{j(\ne i)}^{\cal V}\frac{1}{1+e^{\alpha'_i+\alpha'_j}}=k_i^*,\label{constr-SECM1}\\
\avg{s_i}&&\equiv \sum_{j(\ne i)}^{\cal V}\frac{(\beta_i+\beta_j)^{-1}}{1+e^{\alpha'_i+\alpha'_j}}=s_i^*,\label{constr-SECM2}
\eea
$\forall i$, and by the joint probability distribution
\be
P(\hA,\hW)=\left[\prod_{i<j}^{\cal V}\frac{e^{-(\alpha'_i+\alpha'_j)a_{ij}}}{1+e^{-(\alpha'_i+\alpha'_j)}}\right]
\left[\prod_{i<j}^{{\cal L}}(\beta_i+\beta_j)e^{-(\beta_i+\beta_j)w_{ij}}\right].
\label{P-AW-TS2}
\ee
By definition, in the SECM the parameters defining link probabilities and weights are disentangled, so that the local statistics for these quantities can be set independently. 
In the CECM instead the parameters controlling for link weights also play a role in determining connection probabilities---see eq.~\eqref{ecm_prob}. 
Indeed in the CECM a link $(i,j)$ with high expected weight ($\beta_i+\beta_j\to0$) is forced to be realized ($p_{ij}\to1$), 
and viceversa a link with low expected weight ($\beta_i+\beta_j\to\infty$) becomes unlikely ($p_{ij}\to0$). 
Thanks to the interplay of its parameters, the CECM better captures the dispersion of higher order properties of the network with respect to the SECM, as shown in Fig. \ref{fig:CECM}). 
However, in the CECM connection probabilities, the contribution of parameters $\{\beta_i\}_{i=1}^N$ is logarithmic with respect to that of parameters $\{\alpha_i\}_{i=1}^N$: 
weighted properties in general give only small perturbations to the Lagrange multipliers of node degrees. 
As such, CECM and SECM define similar link probabilities and expected weights (Fig. \ref{fig:CECM}), and are almost interchangeable 
for all practical purposes---the advantage of SECM being an easier numerical implementation. 
Finally, it is noteworthy that CECM and SECM coincide when the constraints on strengths and degrees satisfy $s_i^*=\gamma k_i^*$ $\forall i$ for constant $\gamma$. 
Indeed in this case $\beta_i=\beta_0$ $\forall i$, and thus we have the exact correspondence $\alpha_i'\equiv\alpha_i+\tfrac{1}{2}\log(2\beta_0)$ $\forall i$. 
This is for instance the case of HFBN of Fig. \ref{fig:CECM}.


{\em Final remarks} --- Ensembles of random graphs with given structural properties like those discussed here typically find a twofold application \cite{cimini2018review,squartini2018reconstruction}. 
On one hand, they can be taken as null network models and thus be used to assess the significance of patterns observed for a real network. 
On the other hand, when details on the microscopic structure of a real network are unknown, they can be used to reconstruct the most likely network configuration. 
The grand canonical ensemble introduced here represents, both in its rigorous version and separable approximation, a very versatile tool for these tasks, 
being defined for the most general class of networks with continuous weights. For instance, the {\em fitness-induced configuration model} \cite{cimini2015estimating} 
used to reconstruct networks without degree information is easily implemented in our grand canonical framework \cite{parisi2018pp}.
Extensions of the framework to the case of multiplex networks \cite{bianconi2013statistical} 
as well as to include higher order interactions between generalized coordinates will be covered in future work.

{\em Acknowledgements} --- A.G. acknowledges support from the CNR PNR project CRISIS-Lab. 
R.M, G.C. and G.C. acknowledge support from the EU H2020 projects DOLFINS (grant 640772), SoBigData (grant 654024) and OpenMaker (grant 687941)

\end{document}